\documentclass[]{pasj01}
\draft

\Received{}
\Accepted{}
 
\usepackage{lineno}
\begin{document} 

\title{The radio dichotomy of active galactic nuclei} 

\author{Hubing \textsc{Xiao}\altaffilmark{1}}
\author{Jingtian \textsc{Zhu}\altaffilmark{2,3,6}%
\thanks{Corresponding author}}
\author{Liping \textsc{Fu}\altaffilmark{1}}
\author{Shaohua \textsc{Zhang}\altaffilmark{1}}
\author{Junhui \textsc{Fan},\altaffilmark{2,4,5}%
\thanks{Corresponding author}}

\email{fjh@gzhu.edu.cn}
\email{jingtian.zhu@studenti.unipd.it}

\altaffiltext{1}{Shanghai Key Lab for Astrophysics, Shanghai Normal University, Shanghai, 200234, China}
\altaffiltext{2}{Center for Astrophysics, Guangzhou University, Guangzhou, 510006, China}
\altaffiltext{3}{Department of Physics and Astronomy ``G. Galilei", University of Padova, Padova PD, 35131, Italy}
\altaffiltext{4}{Key Laboratory for Astronomical Observation and Technology of Guangzhou, Guangzhou 510006, China}
\altaffiltext{5}{Astronomy Science and Technology Research Laboratory of Department of Education of Guangdong Province, Guangzhou, 510006, China}
\altaffiltext{6}{Istituto Nazionale di Fisica Nucleare, Padova PD, 35131, Italy}



\maketitle

\begin{abstract}
The question of radio dichotomy in the active galactic nuclei (AGNs) is still in debate even it has been proposed for more than forty years.
In order to solve the old riddle, we collect a sample of AGNs with optical $B$ band and radio 6cm wavelength data to analyze the radio loudness ${\rm log}R$.

Our results indicate a separation of ${\rm log}R = \langle 1.37 \pm 0.02 \rangle$ between radio-loud (RL) AGNs and radio-quiet (RQ) AGNs, suggest the existence of an RL/RQ dichotomy.
For the first time, we suggest combining radio luminosity and radio loudness as a double-criterion to divide AGNs into RLs and RQs to avoid misclassification problems that may happen in the single-criterion scenario, we propose the double-criterion dividing line ${\rm log}L_{\rm 6cm} = -2.7{\rm log}R +44.3$ by using a machine learning method.

In addition, the key point of the RL/RQ dichotomy is the origin of radio emission for the two classes, we suggest the radio emission from RLs and RQs share the same origin, e.g. jets and mini-jets (aborted-jet or outflow), through a correlation study between radio 6cm luminosity and optical $B$ band luminosity.

\end{abstract}


\section{Introduction}
Active galactic nucleus (AGN), one of the most important discoveries of the 20th century, is a compact region at the center of an active galaxy that shows much higher luminosity than the normal galaxy.
AGNs are the most luminous and persistent sources in the Universe and are powered by accretion onto a centered supermassive black hole (SMBH) \citep{Salpeter1964, Lynden-Bell1969}.
The AGN 'zoo' contains various types of AGNs, including quasars, Seyfert galaxies, BL Lacertae objects, Fanaroff-Riley (FR), low-ionization nuclear emission-line region (LINER), etc.
An important property of AGN is the existence of dichotomy between radio-loud (RL) AGN (broad-line radio galaxies, blazars, FR-I, FR-II) and radio-quiet (RQ) AGN (Seyfert galaxies, LINER) \citep{Strittmatter1980, Kellermann1989, White2000ApJS, Balokovic2012, Cao2016, Zhang2021}.

\citet{Strittmatter1980} firstly pointed out that the radio-to-optical flux density ratio (or radio loudness, $R = f_{\rm r}/f_{\rm o}$) for optically selected quasars appears bimodal and suggested quasars can be divided into two distinct populations, namely RL and RQ.
The following studies of radio dichotomy appear controversial.
\citet{Kellermann1989} calculated the radio $R$ based on the 5 GHz flux and 4400 {\AA} flux for 114 bright quasars and claimed the distribution of $R$ appears bimodal; they suggested the RQs with values in the range $0.1 < R <1$, and RLs with $R$ ranging between 10 and 1000.
Later on, \citet{Shastri1993} proposed a value of $R = 10$ to separate RLs and RQs by using $R$ calculated from 5 GHz and optical B (4450 {\AA}) band.
There is more additional evidence from other researchers \citep{Ivezic2002, Zamfir2008, Zhang2021}, and makes the dichotomy seems evident.
However, \citet{White2000ApJS} compiled a large sample of 636 quasars, they found a large population of intermediate radio-loudness sources and claimed there is no evidence in their sample to see a bimodal distribution of radio characteristics and supported by some follow research \citep{Lacy2001, Calderone2013}.
The dichotomy in radio loudness would reveal distinct physical properties of RLs and RQs, such as a different origin of radio emission, different black hole masses, different accretion rates and spins \citep{McLureJarvis2004, Sikora2007, Laor2008, Hamilton2010}.

During the study of the AGN radio dichotomy, we notice that there is a clear sample dependence of dividing value $R$.
\citet{Zhang2021} employed a sample of 2419 sources with optical (B band and V band) and radio (5 GHz) data from the thirteenth quasar and AGN catalogue \citep{Veron2010}.
They suggested a dividing value of ${\rm log} R = 1.26$ to separate AGNs as RLs and RQs, and tested the separation by dividing sources with a statistical dividing line obtaining the maximum separation accuracy in the diagram of radio luminosity against radio loudness.

In this work, we revisit the AGN radio dichotomy with a sample of \citet{Veron2010} catalogue and investigate possible underlie physical properties.
This paper is arranged as follows:
in Section 2, we present our sample and data reduction;
the results are presented in Section 3;
Section 4 will be the discussion of our results;
our conclusion will be presented in Section 5.
The cosmological parameters $H_{\rm 0} = 73 \ {\rm km \cdot s^{-1} \cdot Mpc^{-1}}$, $\Omega_{\rm m} = 0.3$ and $\Omega_{\rm \Lambda} = 0.7$ have been adopted through this paper.

\newpage
\section{The sample}
We compile 2943 sources (2120 quasars, 245 BL Lacs,  578 other AGNs) with available radio 6 cm flux density, apparent magnitudes in $V$ band, colours of $B - V$ from \citet{Veron2010}, and list them in columns (4), (5), (6) of Tab. \ref{fmc}.

Galactic absorption and scattering of electromagnetic radiation by dust and gas between celestial object and observer result in a reddening of the observed spectrum.
We apply a Galaxy extinction by \citet{Schlegel1998}, $E_{\rm B-V}$ (in column (7) of Tab. \ref{fmc}) that is a function of the object position, to obtain the $V$ band magnitude compensation $A_{\rm V}=2.742 \times E_{\rm B-V}$ \citep{Zhang2021}.
A compensated $V$ magnitude is allowed to be translated to $B$ band magnitude with a known colour $B - V$, and further to the flux density $f_{\rm V}$ and $f_{\rm B}$.
Before using both the radio and the optical flux, a $K$-correction $f_{\nu}^{'} = f_{\nu}(1+z)^{\alpha_{\nu}-1}$, in which $f_{\nu}^{'}$ and $f_{\nu}$ are the intrinsic and observed flux densities at frequency $\nu$, $z$ is redshift, $\alpha_{\nu}$ for optical and radio bands are 0.76 and 0.25 from \citet{Zhang2021}, must be employed to obtain the intrinsic intensity from the object.
Radio loudness is defined as the ratio of intrinsic radio flux density to the optical flux density
\begin{equation}
{\rm log} R = {\rm log} \frac{f^{'}_{\rm r}}{f^{'}_{\rm o}},
\label{eq_R}
\end{equation}
where, in this work, we apply radio flux density at 6 cm and optical flux density in $B$ band that are listed in Table \ref{fmc}.
Besides, a monochromatic luminosity is expressed as
\begin{equation}
L_{\rm \nu} = 4 \pi d_{\rm L}^2 \nu f^{'}_{\rm \nu},
\label{eq_L}
\end{equation}
where $d_{\rm L} = (1+z) \cdot \frac{c}{H_{\rm 0}} \cdot \int_{\rm 1}^{\rm 1+z} \frac{1}{\sqrt{\Omega_{\rm M}x^3+1-\Omega_{\rm M}}}dx$ is the luminosity distance.

\newpage
\section{Results}
\subsection{The distribution of sources number versus redshift}
\begin{figure}[h]
\centering
\includegraphics[scale=0.65]{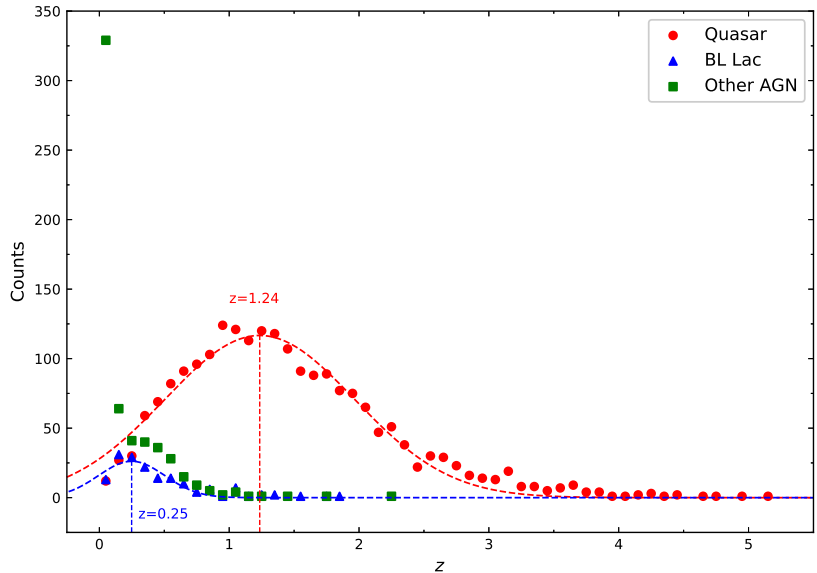}
\caption{The distribution of source number in redshift bins.
The red dot stands for quasar, the blue triangle stands for BL Lac, and the green square stands for other AGN, respectively.
The dashed curves are obtained by fitting data with Gaussian function.}
\label{Fig_z_dis}
\end{figure}

Redshift ranges from 0 to 5.11 for the sources in our sample, except 86 BL Lacs without available redshifts and two radio galaxies (M31 and NGC3031 ) have redshifts of zero.
Fig. \ref{Fig_z_dis} shows the distribution of counts in redshift bins ($bin\_width=0.1$) for all the three classes of sources in our sample.
We manage to fit distributions of quasar and BL Lac use Gaussian functions with $\mu_{\rm Q} = 1.24$, $\sigma_{\rm Q} = 0.07$ and $\mu_{\rm A } = 0.25$, $\sigma_{\rm A} = 0.26$, respectively.
While, we can not apply Gaussian fitting on other AGN distribution because these AGNs in our sample are mostly distributed in the redshift range of 0 to 0.1, as the `green square' shown at the upper-left corner of Fig.\ref{Fig_z_dis}.

\subsection{The radio loudness}
\begin{figure}[h]
\centering
\includegraphics[scale=0.65]{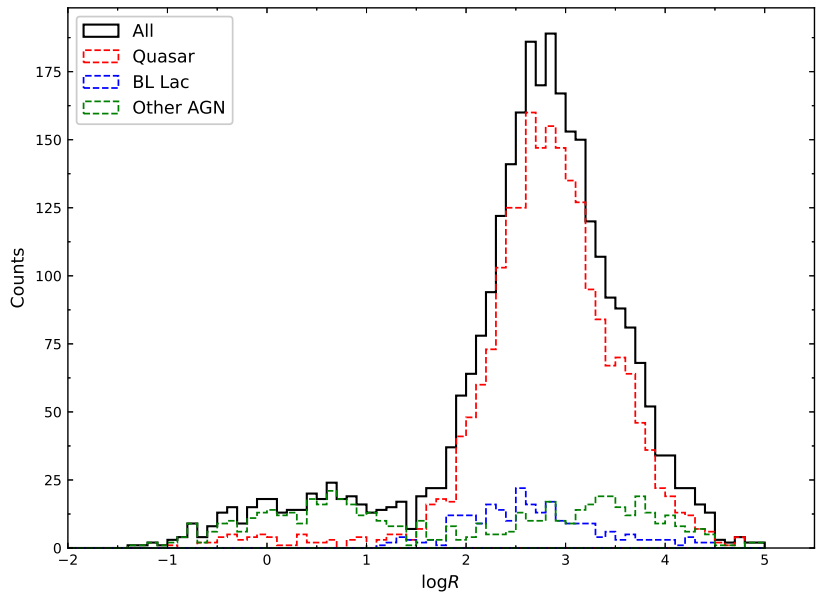}
\caption{The distribution of radio loudness.
The red histogram stands for quasar, the blue one stands for BL Lac, and the green one stands for other AGN, respectively.}
\label{Fig_logR_dis}
\end{figure}

Fig. \ref{Fig_logR_dis} shows the radio loudness (${\rm log} R$) distribution of our sample.
${\rm log} R$ spans more than 6 order of magnitudes for the entire sample, in which quasars have ${\rm log} R$ spans from -0.9 to 5.271 with an average value of $\langle {\rm log} R_{\rm Q} = 2.81 \pm 0.72 \rangle$, BL Lacs have ${\rm log} R$ ranges from 1.131 to 4.473 with an average value of $\langle {\rm log} R_{\rm B} = 2.66 \pm 0.68 \rangle$, and other AGNs ranges from -2.329 to 5.712 with an average value of $\langle {\rm log} R_{\rm A} = 1.89 \pm 1.62 \rangle$.
It is that other AGNs have the widest range and the smallest value of ${\rm log}R$, and quasars have the largest value of ${\rm log}R$ among three classes.

The distribution shows a clear dichotomy, the lower ${\rm log}R$ bump is mainly constructed by other AGNs and some quasars, the higher ${\rm log}R$ bump is mainly constructed by quasars.
There are methods to decompose the distribution into several independent components, a usual way is to fit the whole distribution with a mixture Gaussian model \citep{Fan2016, Zhang2021}.

\begin{figure}[h]
\centering
\includegraphics[scale=0.65]{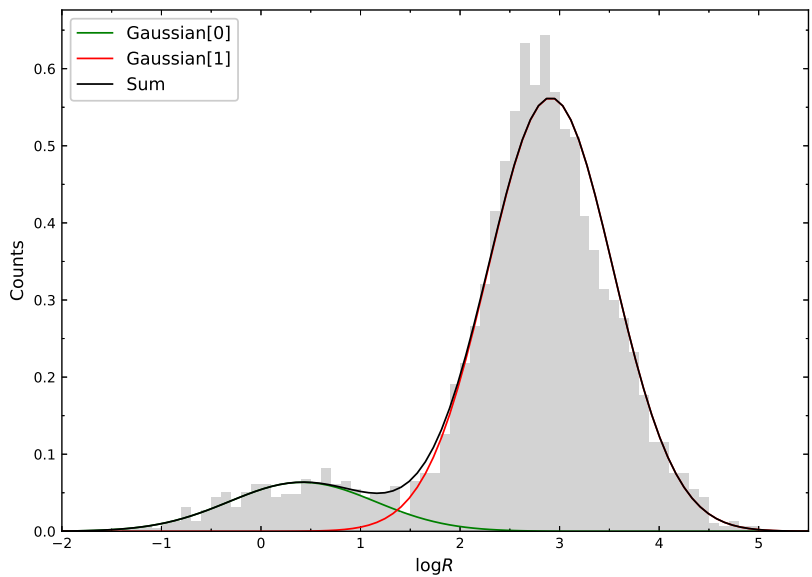}
\caption{The density probability distribution of radio loudness that are fitted with two Gaussian components.}
\label{Fig_logR_gmm_dis}
\end{figure}

Scikit-learn (\textit{sklearn}) is a software machine learning library for the \textit{Python} programming language. 
It features various classification, regression and clustering algorithms.
We apply \textit{Gaussian mixture models} (GMM), which is a probabilistic model that assumes all the data points are generated from a mixture of a finite number of Gaussian distributions with unknown parameters, in the package of \textit{sklearn} to decompose the ${\rm log}R$ distribution as shown in Fig. \ref{Fig_logR_gmm_dis}.
Among the GMM process, we assume there are two \textit{Gaussian} components are structured in the ${\rm log}R$ distribution for the reason that it has been confirmed in previous studies \citep{Shastri1993, White2000ApJS, Zhang2021}.
The proportions of each Gaussian components of the total distribution (the so-called hidden variables) and the mean and standard deviation of each Gaussian components are determined through an \textit{Expectation-Maximization} (EM) algorithm.
This is an iterative algorithm. 
In the `E' step, the initial value of the parameters or the parameters of the last iteration are used to calculate the posterior probability of hidden variables according to the \textit{Bayesian} formula.
In the `M' step, the likelihood function is maximized to obtain the estimated values of the mean and standard deviation.
The `E' and `M' steps are continuously iterated until all the parameters to be estimated converge.
We apply stratified sampling with a scale of 80\% to run the whole GMM process 1000 times in order to increase the generalization ability of our model.
The dichotomy is finally resolved as two Gaussian components in the density probability distribution, the two components give clustering probability of 0.117 for component \textit{Gaussian}[0] and 0.883 for component \textit{Gaussian}[1].
The two components have mean and standard deviation of 
$\mu[0] = \langle 0.42 \pm 0.03 \rangle$ 
and $\sigma[0] = \langle 0.54 \pm 0.03 \rangle$ for curve \textit{Gaussian}[0], 
$\mu[1] = \langle 2.91 \pm 0.01 \rangle$
and $\sigma[1] = \langle 0.39 \pm 0.01 \rangle$ for curve \textit{Gaussian}[1].
The two curves share a cross point at ${\rm log}R = \langle 1.37 \pm 0.02 \rangle$.
In this work, we adopt this cross point as a delimiter to separate radio sources into RLs with ${\rm log}R \geq 1.37$ and RQs with ${\rm log}R < 1.37$.
Subsequently, 2057 quasars, 238 BL Lacs and 314 other AGNs in our sample are classified as RLs, taking 97.0\%, 97.1\% and 54.3\% of their corresponding amounts according to our new classification criteria.
The new classification, denoted to `Radio Class', is listed in column (8) of Tab. \ref{fmc}.

\subsection{The correlation between radio luminosity and optical luminosity}
\begin{figure}[h]
\centering
\includegraphics[scale=0.65]{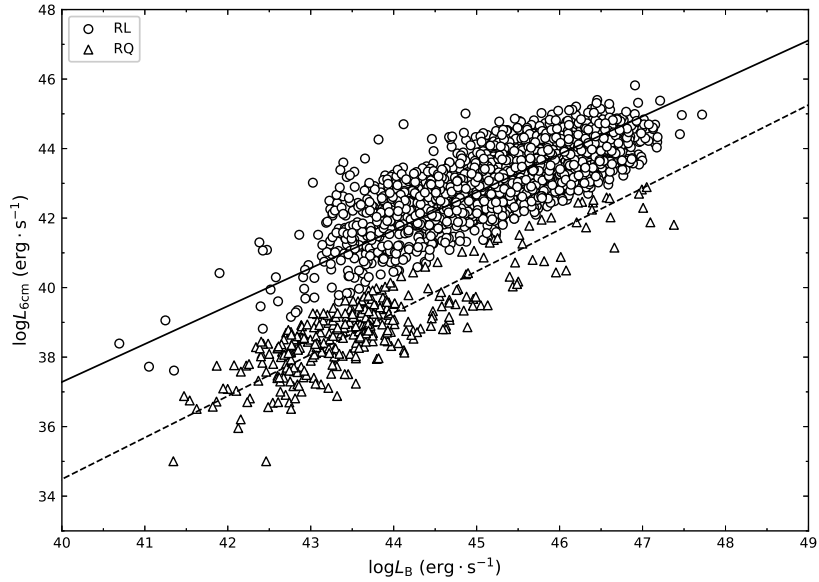}
\caption{The diagram of radio luminosity against B-band luminosity.
The circle and solid line of linear regression for the RLs, the empty triangle and dashed line of linear regression for the RQs.}
\label{Fig_R-O}
\end{figure}

The bimodal ${\rm log}R$ distribution of RLs and RQs imply that the correlation between radio luminosity and optical luminosity could show differences in the two radio classes of sources. 
We analyze the radio luminosity at 6 cm (${\rm log}L_{\rm 6cm}$) and optical luminosity in $B$ band (${\rm log}L_{\rm B}$) with linear regressions, we have
$${\rm log} L_{\rm 6cm} = (1.09 \pm 0.01) {\rm log}L_{\rm B} - (6.40 \pm 0.66),$$
with Pearson partial correlation coefficient $r = 0.34$ and chance probability $p=8.3 \times 10^{-70}$ after removing the redshift effect for RLs; and
$${\rm log} L_{\rm 6cm} = (1.19 \pm 0.03) {\rm log}L_{\rm B} - (13.37 \pm 1.47),$$
with $r = 0.36$ and $p=1.3 \times 10^{-11}$ after removing the redshift effect through Pearson partial analysis for RQs as shown in Fig. \ref{Fig_R-O};

\newpage
\section{Discussion}
\subsection{The source distribution}
We have similar but looser conditions of recruiting sources with respect to the sample selection in \citet{Zhang2021}, so that we have 524 more sources than their sample.
These exceed sources are constructed with 232 quasars, 97 BL Lacs and 196 other AGNs.
Kolmogorov–Smirnov (K-S) test has been employed to investigate the possible difference for redshift, $f_{\rm 6cm}$ and $V_{\rm mag}$ between the sample in two works.
The K-S tests give probabilities of $9.6 \times 10^{-6}$, $1.6 \times 10^{-4}$ and $0.04$ and suggest that the three parameters in this work and that in Zhang's work are not from same distributions.

The distributions of three classes of sources are shown in Fig.\ref{Fig_z_dis}, in which quasars show obviously relative higher mean redshift than BL Lacs and other AGNs (mostly Seyfert galaxies and LINERs).
The BL Lacs and other AGNs are mostly distributed in the range of redshift $0<z<1$, the BL Lac peaks at $z = 0.25$ and other AGNs mostly in $0 < z < 0.1$ redshift bin.
It accords with the observation that BL Lacs, Seyfert galaxies, LINERs are observed in lower redshfits due to their lower optical luminosity with respect to quasars.

Our sample is not complete for the following two reasons.
On one hand, RLs are believed to be 10\% of the entire AGNs \citep{Kellermann2016, Panessa2019}, which means that the quantity of RQs should be much larger than RLs, but there are mostly RLs in our sample.
On the other hand, the redshift distribution of quasars has been reported in \textit{Sloan Digital Sky Survey} (SDSS) studies.
\citet{Paris2018} suggested quasars are mostly ranged from redshift 0 to 3.5 with multi-peaks, but our sample has only one peak at redshift $z=1.24$ for quasars.

The incompleteness of our sample arises from our data selection that we only select the sources with both $B$ band ($V$ band and color index $B-V$) and 6cm wavelength data.
There are large numbers of sources with available $B$ band data, but radio 6 cm observations are only available for some quasars.
Because quasars have stronger radio emission than the rest of the two classes at radio band \citep{Kellermann2016, Radcliffe2021}.
A strong radio emission would require less detecting sensitivity of a radio telescope and a consequently higher probability to be observed.
One should keep in mind that a incompleteness of sample could have some effects on the results of radio dichotomy.
The main influence on our results could be twofold.
(1) Our boundary of ${\rm log}R$ could have been underestimated for the reason of a incomplete sample.
Because the RQ peak should be the dominant component in the bimodal ${\rm log}R$ distribution and spread a much larger range of ${\rm log}R$ than it shows in Fig. \ref{Fig_logR_dis}; 
(2) The correlation between ${\rm log}L_{\rm 6cm}$ and ${\rm log}L_{\rm B}$ could be very different, hard to postulate, in a relatively complete sample. 
A sample with much more RQs than RLs, these RQs could locate at the lower-left region of Fig. \ref{Fig_R-O} in a more complicated way.
Because these RQs are constructed with different types of AGNs that could show different statistic radio-optical relations.

Even the sample is incomplete, we still have the largest sample of studying ${\rm log}R$ bimodal distributions, our results of ${\rm log}R$ shows an agreement with previous studies, that we will discuss in the next section, this agreement should ensure that we are providing reliable study on the subject of the radio dichotomy.


\subsection{The radio dichotomy}
The question of radio dichotomy has been pointed out for more than forty years and studied by many authors with divergent results \citep{Strittmatter1980, Kellermann1989, White2000ApJS, Balokovic2012}.
An important reason of these divergent results is that the previous studies have a small sample size until \citet{Zhang2021}.
They confirmed the existence of the radio dichotomy and claimed a boundary of ${\rm log}R=1.26$ for separating RLs and RQs by using a sample of 2419 radio sources.
Our result of ${\rm log}R= \langle 1.37 \pm 0.02 \rangle$ is different with Zhang's result, the value dispersion of ${\rm log}R$ is caused by different sample size.
The classical boundary value ${\rm log}R=1$, was suggested by \citet{Shastri1993} through analysing a sample of 45 quasars, is clearly smaller than the results of our work and \citet{Zhang2021}.
Through comparison, we notice that ${\rm log}R$ is increasing with the increase of the sample size.
We stress that the radio intensity for RLs should be tens times stronger than the RQs.

There are several interpretations have been proposed for the radio dichotomy \citep{Kellermann1989, Kellermann2016}.
(1) The idea of radio intermittent activity, that suggests quasars may be normally quiescent at radio wavelengths and are observed as radio sources only during the time of unusual activity.
In this scenario, in order to form such a radio dichotomy the variability time scale of at least $10^{5}$ to $10^{6}$ years would be needed.
However, variability time scales are reported typically only in months and years in both RLs and RQs \citep{Barvainis2005}.
(2) Absorption from the intervening plasma local to quasars and synchrotron self-absorption could also be important reasons \citep{Strittmatter1980, Condon1980}.
However, radio searches at the lower band where the opacity is expected to be small do not show any significant change on the RLs to RQs ratio \citep{Ennis1982, Robson1985}.
(3) Geometry effect could also be an important issue of RLs, if we consider the radio emission may be relativistically beamed for the small fraction of quasars with the jet point towards us \citep{ Urry1995, Fan2004, Fan2014, Xiao2015, Pei2016}.
The relativistic beaming models have enjoyed the popularity of interpreting the apparent superluminal motions \citep{Kellermann2004, Xiao2019}.
\citet{Scheuer1979} made predictions about the expected radio detection rate for optically selected quasars, $dN \propto S^{-1-1/p}$, where $N$ is the number density, $p=2+\alpha$ or $3+\alpha$ for different beaming cases.
\citet{Kellermann1989} checked the differential distribution of 6 cm flux density for 75 sources, discovered an excess of sources with $S>30 ~{\rm mJy}$ and claimed the distinction between RLs and RQs can not be caused by the relativistic beaming effect.
(4) Host galaxy is believed to be one of the reasons for radio dichotomy.
\citet{Kellermann2016} have investigated the host galaxies of quasars and found low luminosity $(10^{21} \lesssim L_{\rm 6cm} \lesssim 10^{23} ~ {\rm W/Hz})$ is primarily the result of star formation in the host galaxy.
(5) Stellar or black hole mass, accretion, outflow, or black hole spin are involved to discuss the radio dichotomy by many authors \citep{Wilson1995, Dunlop2003, Ghisellini2004, Sikora2007, Gopal2008, Best2012}
\citet{Laor2008} suggested AGN corona are magnetically heated and that the radio emission in RQs originates in coronal activity when studied the correlation between radio and X-ray flux density and found a robust correlation of $L_{\rm r} \sim 10^{-5}L_{\rm x}$.
\citet{Cao2016} suggested that the radio dichotomy of AGNs predominantly originates from the angular velocity of the circumnuclear gas, RLs with the angular velocity of the circumnuclear gas lower than the critical value at the Bondi radius, while RQs have a larger angular velocity of the circumnuclear gas than the critical value.

Among these interpretations, the former three seem to be expelled.
The host galaxy effect is very promising, but their conclusion is achieved on a sample that is limited by both its size (178 sources) and redshift range ($0.2<z<0.3$).
The last interpretation is model dependent, that involve different ingredients to explain the RL/RQ dichotomy.
However, these models are usually good fit for individual objects.
In a word, the question of radio dichotomy is still open.

Our results show a separation between RL and RQ clusters at the luminosity of \textbf{$L_{\rm 6cm}\sim 10^{40}\ {\rm erg \cdot s^{-1}}$} in Fig. \ref{Fig_R-O}, this separation is similarly mentioned by \citet{Stocke1992}.
This is evidence of a significant difference between the RL and the RQ populations in dependence between their radio and optical powers.
The regressions between ${\rm log}L_{\rm 6cm}$ and ${\rm log}L_{\rm B}$ show strong associations for both the RLs and the RQs, the coefficients stay moderate significant after removing the redshift effect.
We notice that the two slopes are very close to each other, such a close value of slopes of the correlation between ${\rm log}L_{\rm 6cm}$ and ${\rm log}L_{\rm B}$ indicate the radio-optical correlation maybe the same for both the RLs and the RQs.
Therefore, we suggest that the radio emission may originate from the same physical origin for both RLs and RQs. 
The radio emission of RLs is believed to be from the synchrotron radiation of relativistic electrons in the jet.
We, then, suggest a mini-jet or uncollimated outflow exist in the RQs, e.g, the `aborted jet' that intermittently eject material with speed smaller than the escape velocity \citep{Ghisellini2004}. 
The fall back materials collide with the outwards moving matter and generate high energy emission, for instance, X-ray emission.

\subsection{A dividing line between RLs and RQs}
From our study and previous studies, we know that the RL/RQ dichotomy is well separated by the radio loudness.
However, this separation based on the single-criterion of ${\rm log}R$ could cause two kinds of misclassifications:
(1) sources with strong radio luminosity ($L_{\rm r} > 10^{43} \ {\rm erg/s}$) and stronger optical luminosity result in a ${\rm log}R \lesssim 1.37$ refer to RQ class;
(2) source with weak radio luminosity ($L_{\rm r} < 10^{41} \ {\rm erg/s}$) and weaker optical luminosity result in a ${\rm log}R > 1.37$, would be classified as a RL object.
Neither of the two cases is correct, because it is impossible to classify a source with $L_{\rm r} > 10^{43} \ {\rm erg/s}$ to be RQ and $L_{\rm r} < 10^{41} \ {\rm erg/s}$ to be RL. 

\begin{figure}[h]
\centering
\includegraphics[scale=0.65]{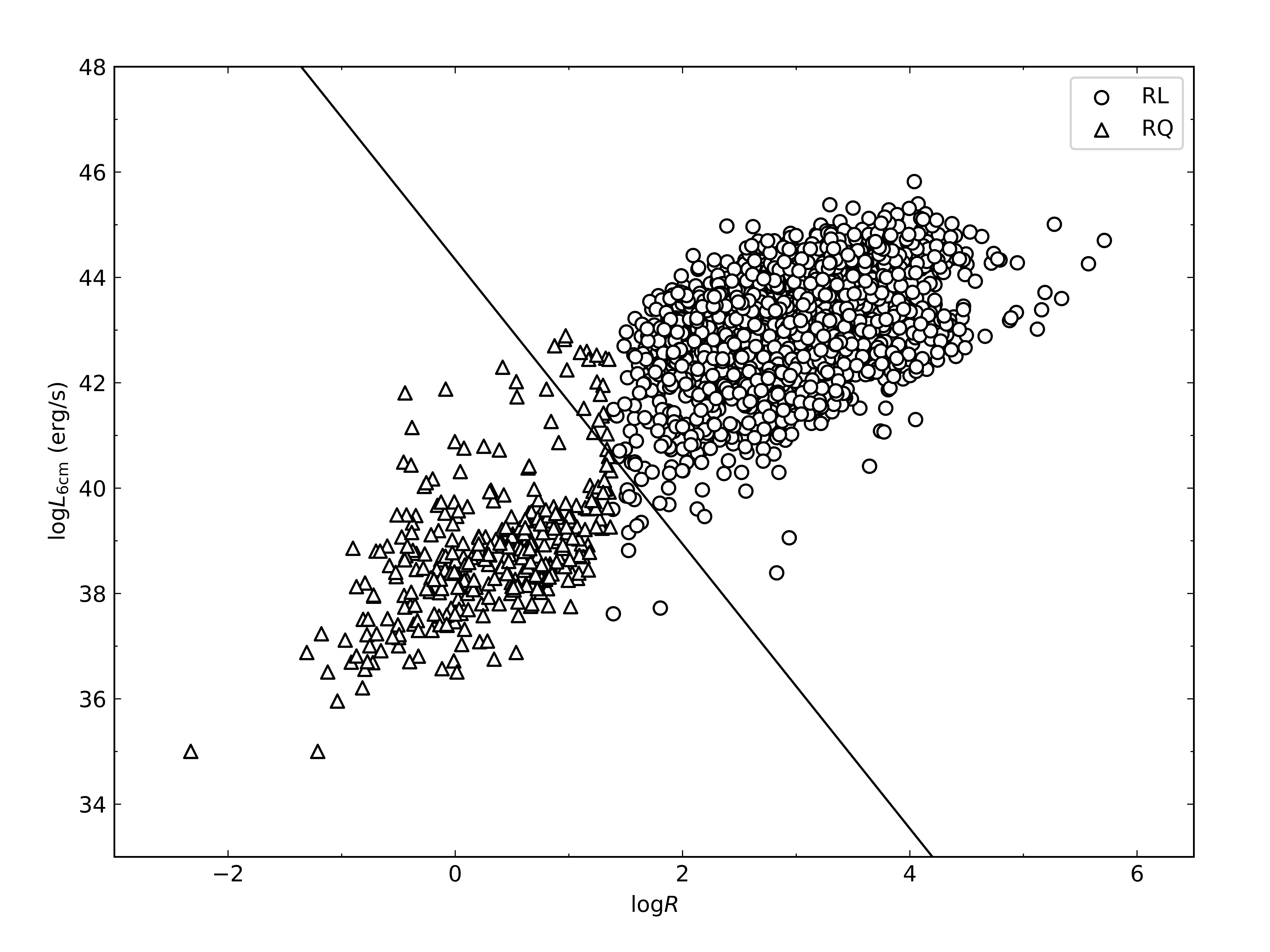}
\caption{The diagram of radio luminosity against radio loudness.
The circle stands for the RLs, the empty triangle stands for RQs, and the black line stands for the dividing line.}
\label{Fig_L_6-logR}
\end{figure}

In order to solve the misclassification problem, we firstly suggest to add one more criterion, which is radio luminosity, to classify RLs and RQs.
Fig. \ref{Fig_L_6-logR} shows the diagram of radio luminosity at 6cm (${\rm log} L_{\rm 6cm}$) against radio loudness (${\rm log} R$).
It is clear that the RLs are occupying the upper-right region and the RQs are taking the lower-left region of the diagram.
Our purpose is to search for a dividing line, that radio luminosity as a function of radio loudness, in the 2D parameters space to separate the RLs and the RQs.
We employ the support vector machine (SVM), a kind of machine learning (ML) method, from \textit{sklearn} to accomplish the task.
SVM is a supervised learning algorithm that is widely used for classification and regression problems.
There are infinite numbers of $N-1$ dimensional hyperplanes can be found to separate two linearly separable samples into different sides of the plane in the $N$ dimensional parameter space.
Then the SVM could make its effort of determining the plane with the maximum margin, i.e., the maximum distance to the nearest samples.
If the two samples are non-linearly separable, SVM can map the samples to a high-dimensional (or even infinite-dimensional) space and find the optimal hyperplane in the high-dimensional space.
The SVM requires a training data set and a data set, that randomly takes 70\% and 30\% sources of each types of RLs and RQs.
The training set is used to find the optimal hyperplane, the test set is used to evaluate the classification accuracy of the optimal hyperplane.
In this work, we put RL and RQ samples in the two-dimensional parameter space, that formed by ${\rm log}R$ and ${\rm log}L_{\rm R}$.
Assuming the hyperplane, is a line in the two-dimensional space, is expressed as $w_{\rm 1}{\rm log}R+w_{\rm 2}{\rm log}L_{\rm R} + m = 0$.
The factors $w_{\rm 1}$, $w_{\rm 2}$ and $m$ can be determined through training SVM with the training set. 
We employ the $svm.LinearSVR$ program provided by $sklearn$ as the SVM classifier.
The hyperparameters of $svm.LinearSVR$ need to be specified before the SVM training starts.
We iterate different combinations of hyperparameters in the SVM training process until $w_{\rm 1}$, $w_{\rm 2}$ and $m$ are found to converge to the value of the maximum margin.
In the end, we get a number of different optimal dividing lines, and the one with the highest accuracy on the test set will be the final optimal dividing line.
The SVM result gives the highest accuracy of 99.0\% for the separation and predicts a dividing line
\begin{equation}
{\rm log} L_{\rm 6cm} = -2.7 \ {\rm log} R + 44.3.
\end{equation}
This dividing line separate our sample into RLs, which occupy the upper right region, and RQs, which take the lower left region of the diagram.
The diagram was also employed in \citet{Zhang2021}, in which they formed a testing line with a slope of -14.4 through analytical method.
The extreme steep dividing line with a high accuracy supported their main results, which is the boundary of ${\rm log}R=1.26$ to separate RLs and RQs.
However, we use this diagram as a tool to solve the misclassification problem, which is arised from single-criterion separation for RLs and RQs.
A much flatter slope of -2.7 for the dividing line through a SVM method in this work.
We suggest a double-criterion of simultaneously using radio loudness (${\log}R=1.37$) and dividing line (${\rm log} L_{\rm 6cm} = -2.7 \ {\rm log} R + 44.3$) to classify radio sources as RLs or RQs. 

\section{Conclusion}
In the present work, we have compiled the largest AGN sample of 2943 sources for the purpose of studying ${\rm log}R$ bimodal distributions, as well as, the radio dichotomy.

We have employed the SVM to explore the distribution of ${\rm log}R$ and managed to decompose it into two Gaussian functions with the GMM.
The value of the two Gaussian curves intersection point, ${\rm log}R = \langle 1.37 \pm 0.02 \rangle$, is applied to be a separation of RLs and RQs.
Consider that the single-criterion of RL/RQ separation could meet misclassifications, we firstly introduce radio luminosity as key ingredients of the separation.
We suggest a double-criterion, ${\log}R=1.37$ and ${\rm log} L_{\rm 6cm} = -2.7 \ {\rm log} R + 44.3$, to classify radio sources as RLs or RQs.

Besides, the correlations between ${\rm log}L_{\rm 6cm}$ and ${\rm log}L_{\rm B}$ show both moderate associations for both RLs and RQs.
And the correlation seem to be the same for both RLs and RQs.
We suggest the radio emission result from a same origin for both RLs and RQs, e.g. jets and mini-jets (aborted-jet or outflow).


\begin{table}
  \tbl{The radio sources of our sample}{%
  \begin{tabular}{lcccccccc}
      \hline
      Name  &  Class  &  Redshift  &  $f_{\rm 6cm}$  &  $V_{\rm mag}$  &  B-V  &  E(B-V)  &  Radio Class  \\
      (1)  &  (2)  &  (3)  &  (4)  &  (5)  &  (6)  &  (7)  &  (8)  \\
      \hline
      SDSS J00037-1108 	&	Q	&	1.57	&	0.118	&	19.93	&	0.33	&	0.0364	&	RL	\\
      FIRST J00051-1010	&	Q	&	1.3	    &	0.043	&	19.21	&	0.42	&	0.0377	&	RL	\\
      UM  18           	&	Q	&	1.9	    &	0.257	&	16.21	&	0.35	&	0.0304	&	RL	\\
      BG CFH 17        	&	Q	&	1.62	&	0.056	&	19.31	&	0.24	&	0.0361	&	RL	\\
      PKS 0003+15      	&	Q	&	0.45	&	0.34	&	16.4	&	0.11	&	0.0491	&	RL	\\
      \hline
    \end{tabular}}
    \label{fmc}
\begin{tabnote}
\footnotemark[$*$] Only five items are displayed.  \\
\end{tabnote}
\end{table}

\begin{ack}
We thank the support from our laboratory, the key laboratory for astrophysics of Shanghai.
Meanwhile, L.P.F acknowledges the support from the National Natural Science Foundation of China (NSFC) grants 11933002, STCSM grants 18590780100, 19590780100, SMEC Innovation Program 2019-01-07-00-02-E00032 and Shuguang Program 19SG41.
S.H.Z acknowledges the support from by Natural Science Foundation of Shanghai (20ZR1473600).
J.H.F acknowledges the support by the NSFC (NSFC U2031201, NSFC 11733001) and the science research grants from the China Manned Space Project with NO. CMS-CSST-2021-A06.
\end{ack}


\begin{thebibliography}{}
\bibitem[Balokovi{\'c} et al. (2012)] {Balokovic2012} Balokovi{\'c}, M., et al. 2012, ApJ, 759, 30.

\bibitem[Barvainis et al. (2005)] {Barvainis2005} Barvainis, R., et al. 2005, ApJ, 618, 108.

\bibitem[Best et al. (2012)] {Best2012} Best, P, N., et al. 2012, MNRAS, 421, 1569.

\bibitem[Calderone et al. (2013)] {Calderone2013} Calderone, G., et al. 2013, MNRAS, 431, 210.

\bibitem[Cao (2016)] {Cao2016} Cao, X. 2016, ApJ, 833, 30.

\bibitem[Condon et al. (1980)] {Condon1980} Condon, J, J., et al. 1980, Nature, 283, 357.

\bibitem[Dunlop et al. (2003)] {Dunlop2003} Dunlop, J, S., et al. 2003, MNRAS, 340, 1095.

\bibitem[Ennis et al. (1982)] {Ennis1982} Ennis, D, J., et al. 1982, ApJ, 262, 460.

\bibitem[Fan et al. (2014)] {Fan2014} Fan, J, H., et al. 2014, RAA, 14, 1135.

\bibitem[Fan et al. (2004)] {Fan2004} Fan, J, H., et al. 2004, ChJAA, 4, 533.

\bibitem[Fan et al. (2016)] {Fan2016} Fan, J, H., et al. 2016, ApJS, 226, 20.

\bibitem[Ghisellini et al. (2004)] {Ghisellini2004} Ghisellini, G., et al. 2004, A\&A, 413, 535.

\bibitem[Gopal-Krishna et al. (2008)] {Gopal2008} Gopal-Krishna, M., et al. 2008, ApJL, 680, L13.

\bibitem[Hamilton et al. (2010)] {Hamilton2010} Hamilton, T, S., et al. 2010, RAA, 407, 2393.

\bibitem[Ivezi{\'c} et al. (2002)] {Ivezic2002} Ivezi{\'c}, {\v{Z}}., et al. 2002, AJ, 124, 2364.

\bibitem[Kellermann et al. (2016)] {Kellermann2016} Kellermann, K, I., et al. 2016, ApJ, 831, 168.

\bibitem[Kellermann et al. (1989)] {Kellermann1989} Kellermann, K, I., et al. 1989, AJ, 98, 1195.

\bibitem[Kellermann et al. (2004)] {Kellermann2004} Kellermann, K, I., et al. 2004, ApJ, 609, 539.

\bibitem[Lacy et al. (2001)] {Lacy2001} Lacy, M., et al. 2001, ApJL, 551, L17.

\bibitem[Laor et al. (2008)] {Laor2008} Laor, A., et al. 2008, MNRAS, 390, 847.

\bibitem[Lynden-Bell (1969)] {Lynden-Bell1969} Lynden-Bell, D. 1969, Nature, 223, 690.

\bibitem[McLure et al. (2004)] {McLureJarvis2004} McLure, R, J., et al. 2004, MNRAS, 353, 45.

\bibitem[Panessa et al. (2019)] {Panessa2019} Panessa, F., et al. 2019, Nature Astronomy, 3, 387.

\bibitem[P{\^a}ris et al. (2018)] {Paris2018} P{\^a}ris, I., et al. 2018, A\&A, 613, 51.

\bibitem[Pei et al. (2016)] {Pei2016} Pei, Z, Y., et al. 2016, Ap\&SS, 361, 237.

\bibitem[Radcliffe et al. (2021)] {Radcliffe2021} Radcliffe, J, F., et al. 2021, A\&A, 649, 9.

\bibitem[Robson et al. (1985)] {Robson1985} Robson, E, I., et al. 1985, MNRAS, 213, 355.

\bibitem[Salpeter et al. (1964)] {Salpeter1964} Salpeter, E, E., et al. 1964, ApJ, 140, 796.

\bibitem[Scheuer et al. (1979)] {Scheuer1979} Scheuer, P, A, G., et al. 1979, Nature, 277, 182.

\bibitem[Schlegel et al. (1998)] {Schlegel1998} Schlegel, D, J., et al. 1998, ApJ, 500, 525.

\bibitem[Shastri et al. (1993)] {Shastri1993} Shastri, P., et al. 1993, ApJ, 410, 29.

\bibitem[Sikora et al. (2007)] {Sikora2007} Sikora, M., et al. 2007, ApJ, 658, 815.

\bibitem[Stocke et al. (1992)] {Stocke1992} Stocke, J, T., et al. 1992, ApJ, 396, 487.

\bibitem[Strittmatter et al. (1980)] {Strittmatter1980} Strittmatter, P, A., et al. 1980, A\&A, 88, 12.

\bibitem[Urry et al. (1995)] {Urry1995} Urry, C, M., et al. 1995, PASP, 107, 803.

\bibitem[V{\'e}ron-Cetty et al. (2010)] {Veron2010} V{\'e}ron-Cetty, M, P., et al. 2010, A\&A, 518, 10.

\bibitem[White et al. (2000)] {White2000ApJS} White, R, L., et al. 2000, ApJS, 126, 133.

\bibitem[Wilson et al. (1995)] {Wilson1995} Wilson, A, S., et al. 1995, ApJ, 438, 62.

\bibitem[Xiao et al. (2019)] {Xiao2019} Xiao, H, B., et al. 2019, SCPMA, 62, 129811.

\bibitem[Xiao et al. (2015)] {Xiao2015} Xiao, H, B., et al. 2015, Ap\&SS, 359, 39.

\bibitem[Zamfir et al. (2008)] {Zamfir2008} Zamfir, S., et al. 2008, MNRAS, 387, 856.

\bibitem[Zhang et al. (2021)] {Zhang2021} Zhang, L, X., et al. 2021, PASJ, 73, 313.

\bibliographystyle{aasjournal}
\end{thebibliography}

\end{document}